\documentclass[12pt]{article}
\usepackage{amssymb}
\usepackage{multirow}
\parskip \baselineskip
\newcommand{\ds}{\displaystyle}

\newcommand{\ol}{\overline}

\newcommand{\ra}{\rightarrow}

\newcommand{\name}[2]{{#1}{\scriptsize{#2}}}
\begin{document}
\begin{center}
{\large \bf Counting or producing all fixed cardinality transversals}

\name{M}{ARCEL} \name{W}{ILD}
\end{center}

\begin{quote}
A{\scriptsize BSTRACT}: An algorithm to count, or alternatively generate, all $k$-element transversals of a set system is presented. For special cases it works in output-linear time.
\end{quote}

\section{Introduction}

Generating {\it all minimal} transversals of a hypergraph ${\cal H}$ based on a set $W$ is a prominent research endeavour [EMG]. But also generating (and evaluating) {\it all} transversals of ${\cal H}$ may be required [W3]. Likewise, the focus may be on {\it all $k$-element} transversals for some integer $k$. For instance in [W2] they need to be counted (not generated) for $k=1$ up to $k=|W|$. As to fixed cardinality constraints in general, see also [BEHM].
While [W2] and [W3] display particular applications of the so called transversal $e$-algorithm, the present paper harks back to [W1] and provides additional theoretic results.

Let us begin with a broader perspective and then zoom in onto transversals. Suppose that $a_1$ up to $a_h$ denote ``constraints'' applying to subsets $X$ of a finite set $W$. Many kinds of combinatorial objects $X$ can be modelled as the sets $X$ that satisfy $h$ suitably chosen constraints. The principle of inclusion-exclusion states that
$$N(a_1 \wedge \cdots \wedge a_h)\quad =\quad 2^w-\ds\sum_{i=1}^h N(\ol{a}_i) + \ds\sum_{1\leq i < j \leq h} N(\ol{a}_i \wedge \ol{a}_j)\quad - \cdots \pm \quad N(\ol{a}_1 \wedge \cdots \wedge \ol{a}_h),$$
where $N(a_1 \wedge \cdots \wedge a_h)$ is the number of $X \subseteq W$ satisfying all constraints, and e.g. $N(\ol{a}_i \wedge \ol{a}_j)$ is the number of $X \subseteq W$ satisfying neither $a_i$ nor $a_j$. Unfortunately $2^h$ terms need to be added or subtracted, and often it is cumbersome to compute the terms  themselves.  

Enter the principle of exclusion (POE) which is discussed in detail in [W1]. Its basic policy  is simply to start with $\mbox{Mod}_0 = 2^W$ and exclude iteratively all sets $X \subseteq W$ that fail to have property $a_1, a_2, \ldots, a_h$. Thus, writing $X \models a_i$ when $X$ satisfies $a_i$, one has:
$$\mbox{Mod}_0\quad \supseteq\quad \mbox{Mod}_1 := \{X \in \mbox{Mod}_0: \ X \models a_1 \}\quad \supseteq\quad \mbox{Mod}_2:= \{X \in \mbox{Mod}_1: \ X \models a_2\}$$
and so forth. Obviously $\mbox{Mod}_h := \{X \in \mbox{Mod}_{h-1} :  \ X \models a_h\}$ comprises exactly the $X$'s that satisfy all constraints, and so $N(a_1 \wedge \cdots \wedge a_h) = |\mbox{Mod}_h|$. This seems like a naive approach but a compact way to pack the members of $\mbox{Mod}_i$ (within so called {\it multivalued rows}) often makes it work. 
In the present article the combinatorial objects at stake are the transversals (or hitting sets) $X$ of a given set system ($=$ hypergraph) ${\cal H} = \{H_1, H_2, \cdots, H_h\}$ of subsets of $W$. Indeed, defining $X \models a_i$ by $X \cap H_i \neq \phi$ unleashes the POE framework.

Here comes the section break up. A medium-size example is given in Section 2, and endowed with theory in Section 4. (Section 3 is discussed in a moment.)
Specifically, transversals can be viewed  as models of a (dual) Horn formula, and hence some facts of [W1] will carry over, but simplify and fortify in the process. This is done in Theorem 4 which exclusively targets fixed cardinality transversals, be it counting or generating. Under quite natural side conditions that can be done in output-linear time. 

Whereas [W1] concentrates on how mentioned {\it multivalued rows} {\it reproduce}, in Section 3 of the present article we focus on {\it individual multivalued rows} $r$ and how the $k$-element sets contained in $r$ can be counted or generated efficiently. We also give the asymptotic number of length $n$ multivalued rows as $n$ goes to $\infty$. Parts of Section 4 depend on Section 3. Section 5 briefly points out the pros and cons of POE as compared to binary decision diagrams.

For positive integers $w$ we put $[w] : = \{1,2, \cdots, w\}$.

\section{The transversal $e$-algorithm by example}

Consider the $(14,6)$-hypergraph with vertex set $W = [14]$ and set ${\cal H} = \{H_1, \cdots, H_6\}$ of hyperedges defined by

\hspace*{2cm} $H_1 = \{3,4,9\}, \quad H_2 = \{5,10\}, \quad H_3 = \{6,7,11, 12\}, \quad H_4 = \{8, 13, 14\},$

\hspace*{2cm} $H_5 = \{1,2,3,4,5,6,7,8\}, \quad H_6 = \{3,4,5,8,12,13\}$.

 As alluded to in the introduction, starting with the powerset $\mbox{Mod}_0:= 2^W$ we filter out the family $\mbox{Mod}_1 \subseteq \mbox{Mod}_0$ of all $X \in \mbox{Mod}_0$ with $X \cap H_1 \neq \emptyset$. Then we filter out the family $\mbox{Mod}_2 \subseteq \mbox{Mod}_1$ of all $X \in \mbox{Mod}_1$ with $X \cap H_2 \neq \emptyset$, and so forth. After having processed $H_h \ (h=6)$, the family $\mbox{Mod}_6$ obviously consists of all transversals of ${\cal H}$.

Under the {\it transversal $e$-algorithm} (or briefly $e$-algorithm) each set $X$ in $\mbox{Mod}_0$ will be identified with its characteristic $0,1$-vector of length $14$. But whenever possible we use the label 2 to indicate that an entry is allowed to be either $0$ or $1$. Thus the powerset is written as  $\mbox{Mod}_0  = (2, 2, 2, 2, 2, 2, 2, 2, 2, 2, 2, 2, 2, 2)$. Actually, it is more precise to write $\mbox{Mod}_0 = \{(2, \cdots, 2)\}$. Similarly set $\mbox{Mod}_1 = \{(2, 2, e, e, 2, 2, 2, 2, e, 2, 2, 2, 2, 2)\}$ because $H_1 = \{3, 4, 9\}$ and a string of symbols $e$ by definition means that only characteristic vectors $X$ are allowed which have {\it at least one} $1$ in a position occupied by an $e$. Similarly we obtain $\mbox{Mod}_2, \mbox{Mod}_3, \mbox{Mod}_4$, but of course we need to introduce subscripts to distinguish the three $e$-constraints. Thus $\mbox{Mod}_4 = \{r\}$ where 
$$r := (2, 2, e_1, e_1, e_2, e_3, e_3, e_4, e_1, e_2, e_3, e_3, e_4, e_4).$$
So far so good, but it's going to be harder to get $\mbox{Mod}_5$ because $H_5$ intersects four $e${\it -bubbles}. As a starter, in view of $H_5\cap H_1 = \{3,4\}$ let us split $r$ into the disjoint union of
$$r[e] := \{X \in r: \ X \cap \{3,4\} \neq \emptyset \} \quad \mbox{and} \quad r[0]  := \{X \in r: \ X \cap \{3,4\} = \emptyset \}.$$
Using our new notation,
$$\begin{array}{lll} r[e] & = & (2, 2, {\bf e}, {\bf e}, e_2, e_3, e_3, e_4, 2, e_2, e_3, e_3, e_4, e_4),\\
\\
r[0] & = & (2, 2, {\bf 0}, {\bf 0}, e_2, e_3, e_3, e_4, 1, e_2, e_3, e_3, e_4, e_4). \end{array}$$
Thus $e_1 e_1 e_1$ is split in $e e 2$ and $0 0 1$.  All $X \in r[e]$ satisfy the fifth constraint since $X \cap H_5 \supseteq X \cap \{3, 4\} \neq \emptyset$. But some $X \in r[0]$ do not satisfy it. In order to exclude these $X$'s and in view of $H_5 \cap H_2 = \{5\}$, we split $r[0]$ into
$$\begin{array}{lll} r[0,e] & := & (2, 2, 0, 0, {\bf 1}, e_3, e_3, e_4, 1, 2, e_3, e_3, e_4, e_4),\\
\\
r[0,0] & := & (2, 2, 0, 0, {\bf 0}, e_3, e_3, e_4, 1, 1, e_3, e_3, e_4, e_4). \end{array}$$
Now all $X \in r[0,e]$ satisfy $X \cap H_5 \neq \emptyset$, but not all $X \in r[0,0]$ satisfy this. Thus, similarly, we split $r[0,0]$ into $r[0,0,e]$  and $r[0,0,0]$. Then $r[0,0,0]$ is split into $r[0,0,0,e]$  and
$$r[0,0,0,0]'\quad =\quad (2, 2, 0, 0, 0, 0, 0, {\bf 0}, 1, 1,  e_3, e_3, e_4, e_4).$$
This row need not be split; the only sets $X \in r[0,0,0,0]'$ satisfying $X \cap H_5 \neq \emptyset$ are the ones with $X \cap \{1, 2\} \neq \emptyset$. They are precisely the elements of
$$r[0,0,0,0]\quad :=\quad (e, e, 0, 0, 0, 0, 0, 0, 1, 1,  e_3, e_3, e_4, e_4).$$
Thus
$$\mbox{Mod}_5\quad =\quad \{r[e], \ r[0,e], \ r[0,0,e], \ r[0,0,0,e], \ r[0,0,0,0]\}.$$

\begin{tabular}{|c|c|c|c|c|c|c|c|c|c|c|c|c|c|l} 
1 & 2 & 3& 4& 5& 6& 7 &8 & 9 & 10 &11 & 12 & 13 & 14 & \\ \hline 
2 & 2 & $e_1$ & $e_1$ & $e_2$ & $e_3$ & $e_3$ & $e_4$ & $e_1$ & $e_2$ & $e_3$ & $e_3$ & $e_4$ & $e_4$ & \qquad $r$ \\ \hline 
\hline
2 & 2 & {\bf e} & {\bf e} & $e_2$ & $e_3$ & $e_3$ & $e_4$ & 2 & $e_2$ & $e_3$ & $e_3$ & $e_4$ & $e_4$ & \qquad $r[e]$ \\ \hline
2 & 2 & {\bf 0} & {\bf 0} & {\bf 1} & $e_3$ & $e_3$ & $e_4$ & 1 & 2 & $e_3$ & $e_3$ & $e_4$ & $e_4$ & \qquad $r[0,e]$\\ \hline
2 & 2& {\bf 0} & {\bf 0} & {\bf 0} & ${\bf e}$ & ${\bf e}$ & $e_4$ & 1 & 1& 2 & 2 & $e_4$ & $e_4$ & \qquad $r[0,0,e]$\\ \hline
2 & 2& {\bf 0} & {\bf 0} & {\bf 0} &{\bf 0} & {\bf 0} & {\bf 1} & 1 & 1 & $e_3$ & $e_3$ & 2 &2 & \qquad $r[0,0,0,e]$\\\hline
$e$ & $e$ & {\bf 0} & {\bf 0} & {\bf 0} & {\bf 0} & {\bf 0} & {\bf 0} & 1 & 1 & $e_3$ & $e_3$ & $e_4$ & $e_4$ & \qquad $r[0,0,0,0]$ \\ \hline \hline
2 & 2& $e$ & $e$ & $e_2$ & $e_3$ & $e_3$ & $e_4$ & 2 & $e_2$ & $e_3$ & $e_3$ & $e_4$ & $e_4$ & \qquad $r_1$\\ \hline
2 & 2 & 0 & 0 & 1 & $e_3$ & $e_3$ & $e_4$ & 1 & 2 & $e_3$ & $e_3$ & $e_4$ & $e_4$ & \qquad $r_2$ \\ \hline
2 & 2& 0 & 0 & 0 & $e_1$ & $e_1$ & ${\bf e}$ & 1 & 1& 2 & 2 & ${\bf e}$ & 2 & \qquad $r_3$\\ \hline
2 & 2 & 0 & 0 & 0 & $e_1$ & $e_1$ & ${\bf 0}$ & 1 & 1 & 2 & 1 & {\bf 0} & 1 & \qquad $r_4$ \\ \hline
2 & 2& 0 & 0 & 0 & 0 & 0 &1 & 1 & 1& $e_3$ & $e_3$ & 2 & 2 & \qquad $r_5$\\ \hline
$e$ & $e$ & 0 & 0 & 0 & 0 & 0  & 0 & 1 & 1 & 2 & {\bf 1} & $e_4$ & $e_4$ & \qquad $r_6$\\ \hline
$e$ & $e$ & 0 & 0 & 0 & 0 & 0 & 0 &1 & 1 & 1 & {\bf 0} & 1 & 2 & \qquad $r_7$ \\ \hline
\end{tabular}

Table 1: Compact representation of a transversal hypergraph

Let us process the rows of $\mbox{Mod}_5$ and sieve out in each row the $X$'s that satisfy $X \cap H_6 \neq \emptyset$. All $X \in r[e]$ satisfy this constraint (because of $e e$ at positions 3, 4), so we carry over $r[e]$ unaltered but relabel it $r_1$. Ditto $r[0,e]$ satisfies the sixth constraint (because of the 1 at position 5) and carries over alias $r_2$. Let $\rho := r[0,0,e]$ and replace $e e$ by $e_1 e_1$ for cosmetic reasons.  Using obvious notation we have $H_6 \cap supp(e_4) = \{8, 13\}$, and so we need to split $\rho$ into
$$\begin{array}{lll} \rho [e] & = & (2, 2, 0, 0, 0, e_1, e_1, {\bf e}, 1, 1, 2, 2, {\bf e}, 2),\\
\\
\rho [0] & = & (2, 2, 0, 0, 0, e_1, e_1, {\bf 0}, 1, 1, 2, 2, {\bf 0}, 1). \end{array}$$
Row $\rho [e]$ carries over alias $r_3$. With obvious notation,  twos$(\rho ) \cap H_6 = \{1, 2, 11, 12\} \cap H_6 \neq \emptyset$, and so row $\rho [0]$ can change and survive as
$$\rho [0;e]\quad =\quad (2, 2, 0, 0, 0, e_1, e_1, 0, 1, 1, 2, {\bf 1}, 0, 1) \quad (=r_4).$$
As to $r[0, 0, 0, e]$, all its members $X$ satisfy $X \cap H_6 \neq \emptyset$  and so $r[0,0,0,e]$ carries over alias $r_5$. But $\sigma := r [0, 0, 0, 0]$ has $H_6 \cap  supp(e_3) = \{12\}$ and needs to be split in

\hspace*{2cm} $\sigma [e]\quad =\quad (e_1, e_1, 0, 0, 0, 0, 0, 0, 1, 1, 2, {\bf 1}, e_4, e_4)$,

\hspace*{2cm} $\sigma [0]\quad =\quad (e_1, e_1, 0, 0, 0, 0, 0, 0, 1, 1, 1, {\bf 0}, e_4, e_4)$.

Row $\sigma [e]$ carries over alias $r_6$, but $\sigma [0]$ in view of $H_6 \cap supp(e_4) = \{13\}$ is further split into

\hspace*{2cm} $\sigma [0,e]\quad =\quad (e_1, e_1, 0, 0, 0, 0, 0, 0, 1, 1, 1, 0, {\bf 1}, 2)$,

\hspace*{2cm} $\sigma [0,0]\quad =\quad (e_1, e_1, 0, 0, 0, 0, 0, 0, 1, 1, 1, 0, {\bf 0}, 1)$.

Row $\sigma [0,e]$ carries over alias $r_7$, but $\sigma [0,0]$ is cancelled since $H_6 \cap X = \emptyset$ for all $X \in \sigma [0,0]$.  To summarize, $\mbox{Mod}_6 := \{r_1, \cdots, r_7\}$ encodes all transversals of the set system ${\cal H}$. 

Due to the disjointness of rows the number $N$ of transversals of ${\cal H}$, i.e. the sum of the cardinalities of the $R=7$ {\it final} rows constituting $\mbox{Mod}_6$, is
$$N=2^3 (2^2-1)(2^2-1)(2^4-1)(2^3-1) + 840 + 288 + 24 + 48 + 18 + 6\quad =\quad 8784.$$
This is fairly evident, and further formalized in Section 4.

\subsection{Another benefit of the $e$-formalism}

This ad hoc subsection fits in well but is not related to the remainder of the paper. Put $W = [w]$. Rather than $\mbox{Mod}_h$ we shall henceforth write $Tr({\cal H})$ for the {\it transversal hypergraph}, i.e. for the family of all transversals of a hypergraph ${\cal H} \subseteq 2^W$. Fixing $A \subseteq W$ we aim to find all $X \in Tr({\cal H})$ with $X \subseteq A$. Dually we may wish to sieve all $X \in Tr({\cal H})$ with $A \subseteq X$. Set ${\cal H}' : = \{H_i \cap A: H_i \in {\cal H}\}$ and ${\cal H}'' : = \{H_i \in {\cal H} : H_i \cap A = \emptyset \}$. Then
$$\{X \in Tr({\cal H}): \ X \subseteq A \} \quad = \quad Tr ({\cal H}')$$
$$\{X \in Tr({\cal H}): \ A \subseteq X \} \ = \ \{A \cup Y: \ Y \in Tr ({\cal H}'')\}$$
Suppose for $1000$ sets $A_j$ one has to solve one of these tasks (or variations thereof). Rather than running the $e$-algorithm $1000$ times for varying ${\cal H}', {\cal H}''$, it's better to run it {\it once} for ${\cal H}$. The $1000$ required set families are then easily obtained from $Tr({\cal H})$. For instance, if ${\cal H} = \{H_1, \cdots, H_6\}$ is as above, then 
$$\{X \in Tr({\cal H}): \ 7 \notin X \ \mbox{and} \  \{8,9\} \subseteq X\}$$
is the disjoint union of these four rows derived from $r_1, r_2, r_3, r_5$ in Table 1: 

\hspace*{1cm} \begin{tabular}{|c|c|c|c|c|c|c|c|c|c|c|c|c|c|} \hline
1 & 2 & 3& 4& 5 &6 & 7 & 8 & 9 & 10 & 11 & 12 & 13 & 14\\ \hline
2 & 2  & $e$ & $e$ & $e_2$ & $e_3$ & ${\bf 0}$ & ${\bf 1}$ & ${\bf 1}$ & $e_2$ & $e_3$ & $e_3$ & 2& 2\\ \hline 
2 & 2 & 0 & 0 &  1  & $e_3$ & ${\bf 0}$ & ${\bf 1}$ & ${\bf 1}$ & 2 & $e_3$ & $e_3$ &2 & 2 \\ \hline 
2 &2 &0 & 0 & 0 &1 & ${\bf 0}$ & ${\bf 1}$ & ${\bf 1}$ & 1 & 2 & 2 & 2 & 2\\ \hline
2 & 2 & 0 & 0 &0 & 0 &  ${\bf 0}$ & ${\bf 1}$ & ${\bf 1}$ & 1 & $e_3$ & $e_3$ & 2 &2 \\ \hline \end{tabular}

\section{Individual $\{0,1,2,e\}$-valued rows}

Here we look at $\{0,1,2,e\}$-valued rows on their own. Thus the row splitting process we glimpsed at in Section 2, and the resulting interdependence of rows, will be postponed to Section 4. Subsection 3.1 gives the formal definition of a $\{0,1,2,e\}$-valued row $r$, along with the number $f(w)$ of such rows of length $w$. In 3.2 and 3.3 we show how the $k$-element sets within a fixed row can be counted, respectively generated. The special case $k=k_{\min}$ deserves extra attention (3.4).

\subsection{Formal definition and number of $\{0,1,2,e\}$-valued rows}

Formally, a $\{0,1,2,e\}$-{\it valued row} on a finite set $W$ is a quadruplet
$$r: = \{\mbox{zeros}(r), \ \mbox{ones}(r), \ \mbox{twos}(r), \ \mbox{ebubbles}(r) \}$$
such that $W$ is a disjoint union of the sets $\mbox{zeros}(r), \cdots, \mbox{ebubbles}(r)$, where any of these may be empty. Furthermore, if ebubbles$(r) \neq \emptyset$ then it is a union of $t \geq 1$ many sets $eb_1, \cdots, eb_t$ (called $e$-{\it bubbles}) such\footnote{It has been observed that a $1$ could be viewed as an $e$-bubble of length one. However, it's better to stick to the given definition and demand a length of at least two. We further note that {\it multivalued} means $\{0,1,2,e\}$-valued in the present article, but can have other meanings in other applications of the POE.} that $\varepsilon_i: = |eb_i| \geq 2$ for all $1 \leq i \leq t$. Thus $r$ can be visualized (up to permutation of the entries) as

(1) \quad $r\quad =\quad (\underbrace{0, \cdots, 0}_\alpha , \underbrace{1, \cdots, 1}_\beta , \underbrace{2, \cdots, 2}_\gamma , \underbrace{e_1, \cdots, e_1}_{\varepsilon_1} , \cdots , \underbrace{e_t, \cdots, e_t}_{\varepsilon_t})$.

By definition, $r$ {\it represents} the family of sets $X \subseteq W$ satisfying

(2) \quad $X \cap \ \mbox{zeros}(r) = \emptyset \ \mbox{and} \ \mbox{ones}(r) \subseteq X \ \mbox{and} \ (\forall 1 \leq i \leq t) \quad eb_i \cap X \neq \emptyset$.

It is however convenient to {\it identify} $r$ with the family of $X$'s satisfying (2). Then, obviously, 

(3) \quad $|r| = 2^\gamma \cdot (2^{\varepsilon_1} -1) \cdots (2^{\varepsilon_t}-1)$.

The Boolean lattice $2^{[w]}$ has $2^{(2^w)}$ many subsets $S$. The $\{0,1,2,e\}$-valued rows of length $w$ yield some of these $S$, but far from all. However, as we shall see, one gets vastly more sets $S$ than with $\{0,1,2\}$-valued rows; the latter merely deliver the $3^w$ many {\it intervals} $S$ of $2^{[w]}$. So let us proceed to calculate the number $f(w)$ of $\{0,1,2,e\}$-valued rows of length $w$. 
Let ${\cal B} \subseteq 2^{[w]}$ be any Boolean sublattice say with bottom and top elements $\perp, \top \in {\cal B}$ and with atoms $A_1, A_2, \cdots, A_s$. Since $A_i \cap A_j = \perp$ for $i \neq j$ and $A_1 \cup A_2 \cup \cdots \cup A_s = \top$, it follows that the sets
$$A_1 \setminus \perp,\quad \cdots \quad , A_\gamma \setminus \perp, \quad A_{\gamma+1} \setminus \perp, \quad \cdots \quad, A_{\gamma+t} \setminus \perp$$
partition $\top \setminus \perp$. Upon permutation we can assume that $s = \gamma + t$ and that for some $\gamma \geq 0$ the sets $A_i \setminus \perp$ are singletons for $i \leq \gamma$, and of higher cardinalities $\varepsilon_1, \cdots, \varepsilon_t$ otherwise. Hence ${\cal B}$ matches a type (1) row $r$ with
$$\begin{array}{rrl}
\mbox{zeros}(r) &= & W \setminus \top, \quad \mbox{ones}(r) \ = \ \perp, \quad
\mbox{twos}(r) \  = \  (A_1 \setminus \perp ) \cup \cdots \cup (A_\gamma \setminus \perp), \\
\\
eb_1 & =& A_{\gamma+1} \setminus \perp, \quad eb_2 \ = \ A_{\gamma +2} \setminus \perp, \quad \mbox{up to} \quad eb_t \ = \ A_{\gamma+t} \setminus \perp. \end{array}$$

Vice versa, every $\{0,1,2,e\}$-valued row $r$ yields\footnote{Notice that $|{\cal B}| < |r|$ but ${\cal B} \not\subseteq r$ for $t> 0$. Of course ${\cal B}=r$ for $t=0$.}  a Boolean sublattice ${\cal B} \subseteq 2^{[w]}$. Thus $f(w)$ equals the number of Boolean sublattices of $2^{[w]}$. As detailed in [IS], this interpretation of $f(w)$ yields
$$f(w)\quad = \quad {\cal B}e \ell \ell (w+2) - {\cal B} e\ell \ell (w+1)$$
where the $n$th Bell number ${\cal B}e \ell \ell (n)$ gives the number of set partitions of a $n$-element set. For instance $f(3) = {\cal B}e \ell \ell (5) - {\cal B} e \ell \ell (4) = 52 - 15 = 37$. Indeed, besides twenty seven $\{0,1,2\}$-valued rows there are three rows of type $(\ast, e, e), (e, \ast, e), (e, e, \ast)$ respectively (where $\ast$ is $0,1,2$), plus the row $(e,e,e)$.

It readily follows from (5.47) in [O] that ${\cal B} e\ell \ell (w+2) - {\cal B} e \ell \ell (w+1)$ is asymptotically equal to ${\cal B} e\ell \ell (w+2)$ as $w \ra \infty$. For all large enough $w$ it e.g. holds that
$$3^w \ll (w^{0.99})^{(w^{0.99})} < {\cal B}e \ell \ell (w+2) < w^w \ll 2^{(2^w)}$$

\subsection{Counting all $k$-element transversals within a row}

In order to calculate the number 
$$\tau_k \quad = \quad \tau_k ({\cal H})$$
 of all $k$-element transversals of a hypergaph ${\cal H}$ on $W$, define
$$\mbox{Card}(r,k) \quad  : = \quad |\{X \in r: \ |X| =k \}|$$
for any $\{0,1,2,e\}$-valued $r$. Obviously $\tau_k$ is the sum of all Card$(r,k)$ where $r$ ranges over all final rows produced by the transversal $e$-algorithm. For $r$ fixed, let us first determine the range of $k$'s for which Card$(r,k) \neq 0$. With notation as in (1) set

(4) \hspace*{1cm} $c_{\min}(r) \quad : = \quad \min \{|X|: \ X \in r\} \quad =  \quad \beta + t.$

Put $X_{\max} = W \setminus \ \mbox{zeros}(r)$. Then $X_{\max}\in r$ and $X \subseteq X_{\max}$ for all $X \in r$, whence
$$c_{\max}(r) \quad : = \quad \max \{|X|: \ X \in r\} \quad = \quad |X_{\max}| \quad = \quad w-\alpha.$$
By (3) it is easy to compute $|r|$, but now we fix $k \in \{c_{\min} (r), \ldots, c_{\max}(r)\}$ and strive for
 Card$(r,k)$. The extreme cases $k_\ast = c_{\min} (r)$ and $k^\ast = c_{\max}(r)$ are trivial:

(5) \quad Card$(r,k_\ast) = \varepsilon_1, \varepsilon_2 \cdots \varepsilon_t$ \ and \ Card$(r,k^\ast) =1$.

Computing Card$(r,k)$ when $k_\ast < k< k^\ast$ is more subtle. It is an exercise (carried out in [W4]) to apply inclusion-exclusion and obtain Card$(r,k)$ as an alternating sum of $2^t$ binomial coefficients. Unless $r$ is long and $t$ is small this method is inferior to the following manner, particularly when Card$(r,k)$ is needed for subsequent values of $k$. We illustrate it on  
$$r_0 \quad : = \quad (e_1, e_1, \ \ e_2, e_2, e_2, \ \ e_3, e_3, e_3, \ \ e_4, e_4, e_4, e_4)$$
and for $w=12$ and $1\leq k \leq 5$:

\begin{tabular}{|c|c|c|c|c|c|} 
$k=$ & 1 & 2& 3 & 4 & 5\\ \hline
& 2 & 1 & 0 &  0 & 0 \\ \hline  
& 0 & 6 & 9 & 5 & 1  \\ \hline
& 0 & 0 & 18 & 45 & 48 \\ \hline 
& 0 & 0 & 0 & 72 & 288\\ \hline
 \end{tabular}
 
 Table 2: Calculating $\tau_k$ recursively

The line $2, 1, 0, 0, 0$ gives the number of sets in $(e_1, e_1)$ having cardinality $1, 2, 3, 4, 5$ respectively. The next line gives the number of sets in $(e_1, e_1, e_2, e_2,  e_2)$ having these cardinalities, and so forth. In general, if $c_1, c_2, \cdots, c_{k-1}$ are the numbers of sets in the segment $(e_1, \cdots,e_1, \cdots, e_{s-1}, \cdots, e_{s-1})$ having cardinality $1,2, \cdots, k-1$ respectively, then the number of sets in the extended segment $(e_1, \cdots, e_1, \cdots, e_{s-1}, \cdots, e_{s-1}, e_s, \cdots, e_s)$ having cardinality $k$ equals

(6) \quad $\ds{\varepsilon_s \choose 1} c_{k-1} + {\varepsilon_s \choose 2} c_{k-2} + \cdots + {\varepsilon_s \choose \varepsilon_s} c_{k-\varepsilon_s}$

This also holds for $k \leq \varepsilon_s$ provided we put $c_i :=0$ for $i \leq 0$. For instance, if we take $s=3$ and $k =5$ in $r_0$, then (6) evaluates to
$${3 \choose 1} c_4 + {3 \choose 2} c_3 + {3 \choose 3} c_2\quad = \quad 3 \cdot 5 + 3 \cdot 9 + 1 \cdot 6 \ = \ 48.$$
As to the calculation of binomial coefficients of type ${\varepsilon \choose 1}, {\varepsilon \choose 2}, \cdots, {\varepsilon \choose \varepsilon}$, they are conveniently calculated as follows:
$${ \varepsilon \choose 1} = \varepsilon, \quad {\varepsilon \choose j+1} = {\varepsilon \choose j} \ \frac{\varepsilon -j}{j+1} \quad \mbox{for} \quad 1 \leq j \leq \varepsilon -1.$$
By {\it first} multiplying with $\varepsilon -j$ and {\it then} dividing by $j+1$ one stays in the realm of integers. Doing this for $\varepsilon = \varepsilon_1$ up to $\varepsilon = \varepsilon_s$  requires $(\varepsilon_1 -1 ) + \cdots  + (\varepsilon_s -1)< w$ multiplications and just as many integer-valued divisions. Applying the $O(w \log w \log\log w) = O(w \log^2w)$ (for shortness) Sch\"{o}nhage-Strassen algorithm for multiplying two $w$-digit numbers (see Wikipedia), the at most $w$ many required binomial coefficients  can be readied in time $O(w^2 \log^2w)$, and they occupy space $O(w^2)$.

\begin{tabular}{|l|} \hline \\
{\bf Theorem 1}: Let $r$ be a $\{0,1,2,e\}$-valued row of length $w$ and let $K \leq w$.\\
Then it costs space $O(w^2)$ and time $O(Kw^2\log^2w)$ to compute the $K$ numbers Card$(r,1)$ \\
up to Card$(r,K)$.\\
\\ \hline
\end{tabular}

{\it Proof.} We assume that $r$ consists only of $t$ many $e$-bubbles, so $\alpha = \beta = \gamma=0$ in (1). Other choices of $\alpha, \beta, \gamma$ only cause trivial adaptions. 
As seen, preparing the binomial coefficients occuring in (6) costs $O(w^2\log^2w)$. For fixed $s \leq t$ consider an initial segment of $e$-bubbles $(e_1, \cdots, e_1, \cdots, e_s, \cdots e_s)$ of lengths $\varepsilon_1, \cdots, \varepsilon_s$ respectively. If Card$'(r,k)$ is the number of $k$-element sets represented by this segment then, as seen in (6), calculating Card$'(r,k)$ involves $\varepsilon_s$ many multiplications of pairs of previously determined at most $w$-digit numbers (and $\varepsilon_s-1$ free additions), whence costs $O(\varepsilon_sw\log^2w)$. Doing this for $1 \leq k \leq K$ gives $O(K\varepsilon_s w\log^2w)$. Summing up yields
$O(K \varepsilon_1 w\log^2w) + \cdots + O(K\varepsilon_t w\log^2w) = O(Kw^2\log^2w)$.

\hfill $\blacksquare$

It is easy to see that the described method to calculate Card$(r,k)$ amounts\footnote{The author adopted this polynomial point of view and the matching Mathematica command {\tt Expand}$[\cdots]$ to get the numbers Card$(r,k)$. Whatever the underlying method of {\tt Expand}$[\cdots]$, for our small values of $w$ that {\it hardwired} command likely beats a {\it high level} Mathematica implementation of the $O(w^2\log^2 w)$ method from Theorem 1.} to expanding a product of some obvious polynomials associated to the $e$-bubbles of $r$. For $r_0$ this gives
$$\begin{array}{lll} 
& & (2x+x^2) \ (3x+3x^2+x^3)^2 \ (4x+6x^2+4x^3+x^4)\\
\\
& = & 72x^4+ 288x^5+534x^6+594x^7+431x^8+208x^9+ 65x^{10}+12x^{11}+x^{12}. \end{array}$$
Here $\mbox{Card}(r_0,4) = \mbox{Card}(r_0, k_\ast) = \varepsilon_1 \varepsilon_2 \varepsilon_3 \varepsilon_4=72$ and Card$(r_0, 12) = \ \mbox{Card}(r_0, k^\ast) =1$ match (5), and $\mbox{Card}(r_0, 5) = 288$ matches Table 2.

\subsection{Generating all $k$-element transversals within a row}

As to {\it generating} all $k$-element members of a $\{0,1,2,e\}$-valued row $r$, let us look at 
$$r= (2,e_2, e_1, 2, 1, e_2, e_1, 0, e_2)$$
and $k=6$. Similar to before we apply recursion according to the partition 
$$\{5\} = \ \mbox{ones}(r), \quad \{1,4\} = \mbox{twos}(r), \quad \{3,7\} \ (\mbox{for}\ e_1), \qquad \{2,6,9\} \ (\mbox{for}\ e_2).$$
Additionally we employ a {\it last in first out} (LIFO) stack management. Namely, the stack starts out with a single ``root object'' $x= (\{5\}, \{1,4\}, [0,2])$. This is a cryptic command that in the next step $x$ needs to split into four sons whose first components are, respectively, the subsets of $\{1,4\}$ with cardinality between $0$ and $2$ joined to $\{5\}$. Each son's second component is the next block of the partition (here $\{3,7\}$). This gives rise to the height four stack in Fig. 1. Notice that $[1,2]$ rather than $[0,2]$ occurs three times because $e_1 e_1$ (as opposed to 22) forbids the empty set. More subtle, in the bottom object $(\{5\}, \{3,7\}, [2,2])$ the entry $[2,2]$ demands that only $\{3,7\}$ itself may eventually be added to $\{5\}$ (because otherwise the final cardinality $k=6$ cannot be reached).

The philosophy of LIFO being that always only the top record of the stack is treated, the second stack gives rise to the third stack in Fig. 1. Its top object gives rise to the final $k$-sets $\{5, 1, 4, 3, 7, 2\}$, $\{5, 1, 4, 3, 7, 6\}$, $\{5, 1, 4, 3, 7, 9\}$. After the next two new top objects have each given rise to three final $k$-sets, the stack has $(\{5,4\}, \{3,7\}, [1,2])$ as its top object. Splitting it yields the fourth stack in Fig. 1. And so on and so forth.

\begin{tabular}{|c|} \hline
$\{5\}, \{1, 4\}, [0,2]$ \\ \hline \end{tabular} \ $\ra$ \  \begin{tabular}{|c|} \hline
$\{5,1,4\}, \{3,7\}, [1,2]$ \\ \hline
$\{5,4\}, \{3,7\}, [1,2]$ \\ \hline 
$\{5,1\}, \{3,7\}, [1,2]$ \\ \hline
$\{5\}, \{3,7\}, [2,2]$ \\ \hline \end{tabular} \ $\ra$ \ \begin{tabular}{|l|} \hline
$\{5, 1, 4,3,7\}, \{2,6,9\}, [1,1]$\\ \hline
$\{5,1,4,7\}, \{2,6,9\}, [2,2]$\\ \hline
$\{5,1,4,3\}, \{2,6,9\}, [2,2]$\\ \hline
$\{5,4\}, \{3,7\}, [1,2]$ \\ \hline
$\{5,1\}, \{3,7\}, [1,2]$ \\ \hline
$\{5\}, \{3,7\}, [2,2]$ \\ \hline \end{tabular} \ $\ra \cdots$ \

$\ra$ \ \begin{tabular}{|l|} \hline
$\{5,4, 3,7\}, \{2,6,9\}, [2,2]$ \\ \hline
$\{5, 4,7\}, \{2,6,9\}, [3,3]$ \\ \hline
$\{5, 4, 3\}, \{2,6,9\}, [3,3]$\\ \hline
$\{5,1\}, \{3,7\}, [1,2]$ \\ \hline
$\{5\}, \{3,7\}, [2,2]$ \\ \hline \end{tabular} \ $\ra \cdots$

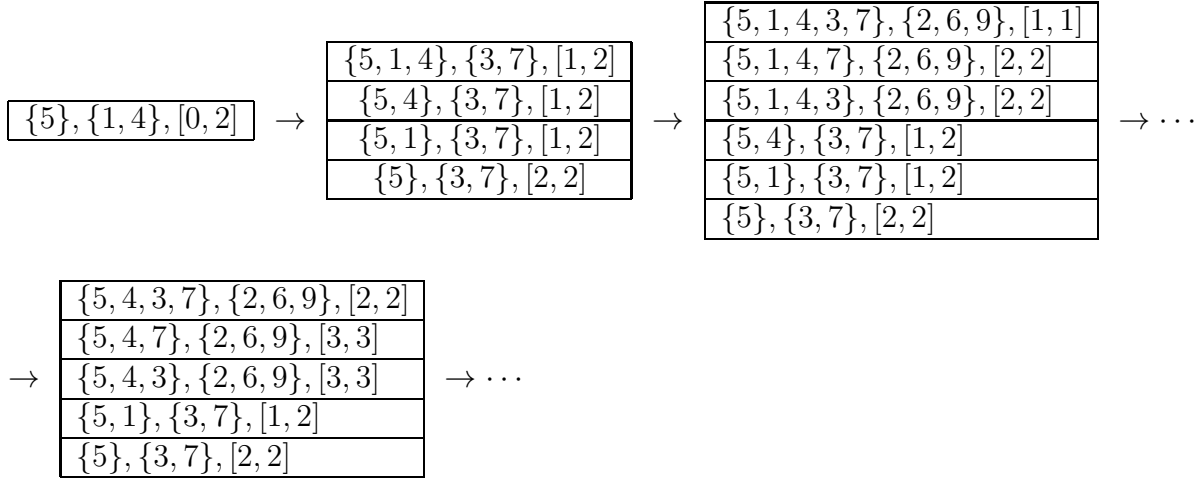
Fig. 1: Generating all $k$-element transversals with LIFO

\begin{tabular}{|l|} \hline \\
{\bf Theorem 2:} Let $r$ be a $\{0,1,2,e\}$-valued row of length $w$ and let $k \leq w$\\
 be fixed. Then
the sets $X \in r$ with $|X| = k$ can be generated in time $O(w^2Card(r,k))$.\\
\\ \hline \end{tabular}

{\it Proof.} We first make precise how the top object $(A,B, [i,j])$ in the sketched LIFO algorithm is to be split. Here $A \subseteq W$ is the accumulated target set, and $B\subseteq W$ is the $e$-bubble to some $e_m e_m \cdots e_m$ (see (1)), and $[i,j]$ by induction is the appropriate subinterval of the integer interval $[1, |B|]$. The ``sons'' of $(A,B, [i,j])$ must be of type $(C,D, [\ast, \ast])$ where $D$ is the $e$-bubble\footnote{For convenience we assume that $m+1, m+2$ are still  $\leq t$. Otherwise special cases arise that are similarly handled.} to $e_{m+1} \cdots e_{m+1}$, and $C$ can be any of the sets $A \cup B'$ where $B'$ ranges over all subsets of $B$ with cardinality between $i$ and $j$. What is the interval $[\ast, \ast ]$ for  a particular fixed $C$? Recalling that $k$ is the final cardinality to be achieved, and putting $\delta : = k - |C|$, a moment's thought shows that
$$[\ast, \ast ]\quad =\quad [\max (1, \delta - \varepsilon_{m+2} - \cdots - \varepsilon_t), \quad \min (\varepsilon_{m+1}, \delta - \sigma)]$$
where $\sigma$ is the cardinality of $\{m+2, m+3, \cdots, t\}$.

Running the LIFO algorithm amounts to building a rooted tree $T$ whose leaves correspond to the $Card(r,k)$ sets $X \in r$ with $|X| = k$. 
The unique path from a leaf $X$ to the root hence traces $t+2$ nodes. For instance:
$$X = \{5,1,4, 3,7,2\} \quad \ra \quad  \{\{5,1,4, 3,7\}, \{2, 6, 9\}, [1,1] ) \quad \ra$$
$$(\{5,1,4\}, \{3,7\}, [1,2]) \quad \ra \quad (\{5\}, \{1,4\}, [0,2]).$$
These nodes correspond to the objects that were split to create $X$. The claim follows from $|T| \leq (t+2)Card(r,k) \leq w \ \mbox{Card}(r, k)$ and the fact that each object in $T$ requires work $O(w)$, as is clear from the above. \hfill $\blacksquare$

It is easy to see that $O(w^2)$ is the maximum size of the LIFO stack  in Fig.1; this height can be much smaller than $Card(r,k)$.

\subsection{The special case $k=k_{\min}$}

The important {\it transversal number} of a set system ${\cal H}$ 
is defined as 
$$k_{\min} ({\cal H})  \quad : = \quad \min \{|X|: \ X \in {\cal T}r({\cal H}) \}$$
For instance, finding the minimum number of pieces necessary in a set covering problem amounts to determine $k_{\min} = k_{\min} ({\cal H})$ for some associated hypergraph ${\cal H}$. Note that $k_{\min}$ {\it as well as} $\tau_{\min}:= \tau_{k_{\min}}$  can be gleaned at once from a representation of ${\cal T}r({\cal H})$ by $\{0,1,2,e\}$-valued rows. For instance, with respect to Table 1 we get from (4) that:
$$\begin{array}{lll}
k_{\min} & =& \min \{c_{\min}(r_1), \cdots, c_{\min} (r_7) \} \\
\\
&  = & \min \{0+4, \ 2+2, \ 2+2, \ 4+1, \ 3+1, \ 3+2, \ 4+1 \} \quad = \quad 4. \end{array}$$
Using (5) that gives 
$$\begin{array}{lll} \tau_{\min} & = &\tau_4 \ = \ \mbox{Card}(r_1,4) + \ \mbox{Card}(r_2, 4) + \ \mbox{Card}(r_3,4) + \ \mbox{Card}(r_5,4) \\
\\
& = & (2 \cdot 2 \cdot 4 \cdot 3) + (4 \cdot 3) + (2 \cdot 2) + 2  \  =\ 66. \end{array}$$
It is evident that also {\it generating} all transversals $X$ with $|X| = k_{\min}$ can be done more smoothly than in Section 3.3.  
The minimum-cardinality transversals constitute a sub-family of the popular [EMG] inclusion-minimal transversals. The $e$-algorithm seems to be predestined to handle that subfamily, although it isn't easy to formally assess its performance (work in progress).

\section{The transversal $e$-algorithm in theory}

If a seventh constraint corresponding to say $H_7 = \{3,4,5\}$ were to be imposed in Table 1, this would cause the cancellation of $r_3$ to $r_7$, and so the work to produce these (multivalued) rows would have been in vain. Fortunately such costly {\it deletions of rows can be prevented} by looking ahead. Specifically, any POE-produced row is called {\it feasible} if it contains at least one model $X_0$.  
Because $r$ is the disjoint union of its ``candidate sons'' $r[e], r[0,e], r[0,0,e]$ and so forth (Section 2), at least one of them will remain feasible. As opposed to other applications of the POE, here feasibility is easily tested. Namely, $\ol{r}$ is feasible if and only if

(7) \qquad $(\forall 1 \leq i \leq h) \quad H_i \not\subseteq \ \mbox{zeros} (\ol{r})$.

Obviously (7) is necessary, and it is sufficient because then $X_{\max} = W \setminus \mbox{zeros}(\ol{r})$ is a model. The non-feasible sons can hence be deleted right away. More generally, fix $k \in [w]$ and call $r$ {\it extra feasible} if it contains a model of cardinality $\geq k$. 
The above remarks constitute the essence of the proof of Theorem 3. 

\begin{tabular}{|l|} \hline \\
{\bf Theorem 3}:  Let ${\cal H}$ be a $(w,h)$-hypergraph, and let 
$k \in [w]$.
Then the\\
transversal $e$-algorithm can be adapted to calculate: \\ \\
 a) The number $N$ of all transversals of ${\cal H}$ in time $O(Nh^2w^2)$;\\ \\
 b) The number of $N$ of all at least $k$-element transversals of ${\cal H}$ in time $O(Nkh^2w^2\log^2w)$.\\ \\ \hline \end{tabular}

{\it Proof.} As before we think of $r_0=(2,2,\cdots, 2)$, with components labelled by the elements of $W=[w]$, as the powerset of $W$. Initially the ``working stack'' solely comprises the row $r_0$ with  the pointer $PC(r_0) =1$ (where $PC$ stands for pending constraint). Note that $r_0$ is extra feasible since $W \in r_0$. Generally, the top row $r$ of the working stack is treated as follows. If $PC(r) = j$ (for some $j \in [h]$) then the hyperedge $H_j \in {\cal H}$ is ``imposed'' upon $r$, 
which means that the set $U$ of all $X \in r$ with $X \cap H_j \neq \emptyset$ is represented as a disjoint union of $s \leq w$ many rows $r_1, \cdots, r_s$. According to [W1, Section 5], this is always possible. (Section 2 of the present article illustrates the most subtle case.) Writing $U$ as $r_1 \cup r_2 \cup \cdots \cup r_s$ costs $O(sw) = O(w^2)$. Because $r$ was extra feasible by induction, at least one of its candidate sons $r_j$ will be as well. Since the extra feasibility of $r_j$ amounts to the truth of both (7) and $|X_{\max}|\geq k$, it costs $O(shw) = O(hw^2)$ to sieve the {\it sons} of $r$, i.e. the extra feasible rows amoung $r_1, \cdots, r_s$. Altogether the cost of one imposition of a constraint upon a  row is $O(w^2) + O(hw^2)= O(hw^2)$.

The $R$ final rows can be viewed as the leaves of a tree with root $(2,2,\cdots 2)$ that has height $h$; each imposition triggers all sons of some node. Therefore the number of impositions is at most $Rh$ (distinct final rows possibly having some of their $h$ forfathers coinciding).  It follows that producing the $R$ final rows costs $O(Rh \cdot hw^2) = O(Nh^2w^2)$ in view of $R \leq N$, by the disjointness of final rows. Counting all transversals within a row costs $O(w)$ by (3), whence doing it for all rows costs $O(Nw) = O(Nh^2w^2)$. This yields claim (a).

As to (b), by Theorem 1 it costs $O(kw^2\log^2w)$ to count the
$$|r| - \ \mbox{Card}(r,1) - \ \mbox{Card}(r,2) - \cdots -  \ \mbox{Card}(r,k-1)$$
many transversals $X \in r$ with $|X| \geq k$. Doing it for all final rows costs $O(Nkw^2\log^2w)$. Claim (b) thus follows from

\hspace*{4cm} $O(Nh^2w^2) + O(Nkw^2\log^2w) = O(Nkh^2w^2\log^2w)$. \hfill $\square$

As is clear from the proof, the $O(Nkh^2w^2\log^2w)$ bound can be improved to $O(Rkh^2w^2\log^2w)$ where $R \leq N$ is the mentioned number of final $\{0,1,2,e\}$-valued rows. Albeit in practise $R$ is often much smaller than $N$, the only obvious theoretic upper bound of $R$ is $N$. If rather than counting we must\footnote{In practise, generating all of them is mainly necessary for exact optimization, but then one rather generates them {\it bunch-wise} in multivalued rows.} {\it generate} all relevant transversals one by one, then we have no choice between $R$ and $N$ but are stuck with the latter.

Let $s_{\max}$ be the maximum number of sons of a multivalued row that occurs in any fixed run of the POE (whether $e$-algorithm or something else). According to [W1, Thm.6] using a LIFO stack management (akin to Section 3.3) reduces the {\it space} requirement of POE-counting to $O (hws_{\max})$. It is easy to see that for the $e$-algorithm one has $s_{\max} \leq \min \{d, \frac{w}{2}\}$ where $d: = \max \{|H_i|: \ 1 \leq i \leq h\}$, and so $O(hws_{\max}) = O(hw^2)$ is independent of $N$. 

Notice that $X$ is a transversal of $H_1, \cdots, H_h$ if and only if its complement $X^c = W \setminus X$ is a {\it noncover} in the sense that $X^c \not\supseteq H_i$ for all $1 \leq i \leq h$. Although the $e$-algorithm can thus count (or generate) noncovers, it pays to introduce the symbolism $n n \cdots n:=$ ``at least one 0'' and a corresponding {\it noncover $n$-algorithm} which produces the noncovers ``directly'', not as $X^c$. The noncover $n$-algorithm in turn generalizes to the {\it Horn $n$-algorithm} of [W1] which counts the models of any given Horn formula. Because Theorem 3a and Theorem 3b above correspond to not so obvious special (and dualized) cases of [W1, Thm.2] respectively [W1, Thm.7], we deemed it worthwile to offer a fresh proof. Even more so because (7) is much smoother than the corresponding feasibility test for general Horn formulae. Theorem 4 below transfers further results of [W1] about {\it fixed} cardinality models to our framework. Its proof is omitted (being along the lines of the proof above) but we mention that Theorem 1 and Theorem 2 are used throughout. They appeared already as statements (16) and (15) in [W1], but their proofs were postponed\footnote{The $O(Kw^2\log^2w)$ bound in Theorem 1 actually improves upon the $O(Kw^3)$ bound in [W1, (16)]. This entails that $(a')$ in Theorem 4 above could be slightly improved accordingly; we omitted it in order to minimize confusion.} to the present article.

\begin{tabular}{|l|} \hline \\
{\bf Theorem 4:} Let ${\cal H}$ be a $(w,h)$-hypergraph and let $k \in [w]$. To avoid trivial\\
 special cases we assume that the number $N$ of various models considered below, is $> 0$.\\
  Define $R \leq N$ as the number of final rows delivered by the transversal $e$-algorithm \\
  when applied to ${\cal H}$.\\
\\
(a) [W1, Thm.10] The number $N$ of transversals of ${\cal H}$ with $|X| = k$ can be calculated in \\
\hspace*{.5cm} time $O(R2^hh w^4k)$.\\
\\
$(a')$ [W1, remark to Thm.10] The $N$ transversals of ${\cal H}$ with $|X| =k$ can be generated in \\
\hspace*{.6cm} time $O(N2^hhw^5)$.\\
\\
(b) [W1, Thm.8] Suppose that $h \leq k \leq w$. Then the number $N$ of transversals $X$ of ${\cal H}$\\ 
\hspace*{.5cm} with $|X| =k$ can be calculated in time $O(Rkh^2w^3)$. \\
\\
$(b')$ [W1, Thm.4] Suppose that $h \leq k \leq w$. Then the $N$ transversals of ${\cal H}$ with $|X| =k$ \\
\hspace*{.6cm} can be generated in time $O(Nh^2w^2)$.\\ 
\\
(c) [W1, Thm.9] Suppose the number of $k'$-element transversals increases as $k'$ ranges \\
 \hspace*{.5cm} from $w$ down to $k$. Then the number $N$ of ${\cal H}$-transversals $X$ with $|X| =k$ can \\
\hspace*{.5cm} be calculated  in time $O(Nh^2w^5)$.\\
\\
 \hline \end{tabular}

\section{Conclusion} 

In [W4], which is a somewhat  verbose preliminary version of the present article, a Mathematica implementation of the $e$-algorithm is pitted against Mathematica implementations of (a) inclusion-exclusion,  (b) lexicographic generation, and (c) the ``hardwired'' whence advantaged Mathematica command {\tt SatisfiabilityCount}. The latter is based on binary decision diagrams (BDD's).

Broadly speaking, the  $e$-algorithm combines the advantages of inclusion-exclusion and {\tt SatisfiabilityCount} without adopting their disadvantages. Let $\tau$ be the number of all transversals.  The advantage of inclusion-exclusion is that calculating all $\tau_k\, (1\leq k\leq w)$ doesn't take much longer than calculating $\tau$ (for fixed $h$ time scales about proportional to $w$), its disadvantage the ominous factor $2^h$. The advantage of {\tt SatisfiabilityCount} is its benign exponential dependence on $h$. Its disadvantage is the inability of BDD's to handle fixed-cardinality constraints.  

Albeit some of the experimential results in [W4] remain interesting, the author also accepts the following criticism of one Referee: 
\begin{quote}
{\tt SatisfiabilityCount} is a function to count the solutions of a satisfiability problem, and transversals are only a special case, so the function is ``abused'' (in particular when lots of artificial constraints are added to find transversals of a certain size!) to perform a task it was not programmed for.
\end{quote}

But then again, the principle of exclusion (Section 1) continues to tease {\tt SatisfiabilityCount} when the issue is counting (let alone generating) the models of an {\it arbitrary} Boolean function in CNF, provided it happens to have few or no models. This is work in progress, and so are other applications of POE. If Mathematica code algorithms compare favorably with corresponding hardwired Mathematica commands, obviously the former algorithms are inherently superior. It has been suggested (fairly or not) that Mathematica commands aren't state of the art, and hence the author's POE-algorithms should be implemented in $C^+$ (say) and compared to existing $C^+$-implementations. Being not familiar with $C^+$ (and too lazy to learn), I leave that worthwile task to others. 

See also Section 9 in [W1] for further analysis of the pros and cons of POE.

\section*{Acknowledgement.} I am grateful to Stephan Wagner and Andrew Odlyzko for pointing out references [IS] respectively [O].

\vskip 1.5cm

\section*{References}
\begin{enumerate}
\item[{[BEHM]}] M. Bruglieri, M. Ehrgott, H.W. Hamacher, F. Maffioli, An annotated bibliography of combinatorial optimization problems with fixed cardinality constraints, Disc. Appl. Math. 154 (2006) 1344-1357.
 \item[{[EMG]}] T. Eiter, K. Makino, G. Gottlob, Computational aspects of monotone dualization: A brief survey, Discrete Appl. Math. 156 (2008) 2035-2049.
 \item[{[IS]}] The Integer-Sequences-Webpage, http://oeis.org/A005493.
 \item[{[O]}] A. Odlyzko, Asymptotic enumeration methods, Handbook of Combinatorics Vol.2, 1063-1229, Elsevier 1995.
\item[{[W1]}] M. Wild, Compactly generating all satisfying truth assignments of a Horn formula, Journal on Satisfiability, Boolean Modeling and Computation 8 (2012) 63-82. 
\item[{[W2]}] M. Wild, S. Janson, S. Wagner, D. Laurie, Coupons collecting and transversals of hypergraphs, to appear in DMTCS.
\item[{[W3]}] M. Wild, Computing the output distribution and selection probabilities of a stack filter from the DNF of its positive Boolean function, Journal of Math. Imaging and Vision, online, 1 August 2012.
\item[{[W4]}] M. Wild, Counting or producing all fixed cardinality transversals, preliminary version of the present article, arXiv : 1106.0141v1.
\end{enumerate}

\end{document}